\documentclass[aps,prc,amsmath,reprint,superscriptaddress,longbibliography]{revtex4-2}

\usepackage{amsfonts}
\usepackage{dcolumn}
\usepackage{graphicx}
\usepackage{xcolor}
\usepackage{subfig}

\newcommand\muminus{\ensuremath{\mu^{-}}}

\newcommand\MeV{\ensuremath{\text{~MeV}}}
\newcommand\MeVc{\ensuremath{\text{~MeV}/c}}
\newcommand\keVc{\ensuremath{\text{~keV}/c}}

\newcommand\um{\ensuremath{\text{~}\mu\text{m}}}
\newcommand\us{\ensuremath{\text{~}\mu\text{s}}}

\begin{document}

\title{Charged particle spectra from $\muminus$ capture on Al}

\author{A.~Gaponenko}
\email{gandr@fnal.gov}
\affiliation{Fermi National Accelerator Laboratory, Batavia, Illinois 60510, USA.}

\author{A.~Grossheim}
\affiliation{TRIUMF, Vancouver, British Columbia, V6T 2A3, Canada}

\author{A.~Hillairet}
\affiliation{University of Victoria, Victoria, British Columbia, Canada.}

\author{G.M.~Marshall}
\affiliation{TRIUMF, Vancouver, British Columbia, V6T 2A3, Canada}

\author{R.E.~Mischke}
\affiliation{TRIUMF, Vancouver, British Columbia, V6T 2A3, Canada}

\author{A.~Olin}
\altaffiliation[Affiliated with: ]{University of Victoria, Victoria, British Columbia, Canada.}
\affiliation{TRIUMF, Vancouver, British Columbia, V6T 2A3, Canada}

\date{\today}

\begin{abstract}
Published data on the emission of charged particles following nuclear
muon capture are extremely limited.  In addition to its interest as a
probe of the nuclear response,
these
data are important for the design of some current searches for lepton
flavor violation.  This work presents momentum spectra of protons and
deuterons following $\muminus$ capture in aluminum.  It is the first
measurement of a muon capture process performed with a tracking
spectrometer.
A precision of better than 10\% over the momentum range of
100--190\MeVc{} for protons is obtained; for deuterons of
145--250\MeVc{} the precision is better than 20\%.
The observed partial yield of protons with emission momenta above
80\MeVc{} (kinetic energy 3.4\MeV) is
$0.0322\pm0.0007(\text{stat})\pm0.0022(\text{syst})$
per capture, and for
deuterons above 130\MeVc{} (4.5\MeV) it is
$0.0122\pm0.0009(\text{stat})\pm0.0006(\text{syst})$.
Extrapolating to total yields gives
$0.045\pm0.001(\text{stat})\pm0.003(\text{syst}) \pm 0.001(\text{extrapolation})$
per capture for protons and
$0.018\pm0.001(\text{stat})\pm0.001(\text{syst})\pm 0.002(\text{extrapolation})$
for deuterons, which are the most precise
measurements of these quantities to date.

\end{abstract}

% \pacs{23.40.-s, 25.30.Mr}
%\keywords{}

\maketitle

%%%%%%%%%%%%%%%%%%%%%%%%%%%%%%%%%%%%%%%%%%%%%%%%%%%%%%%%%%%%%%%%

\section{Introduction}

When a negative muon loses energy by ionization and scattering in
matter it encounters the field of the nucleus and at kinetic
energies below the order of 10--100 eV it can undergo
\textit{atomic} muon capture \cite{Haff:1974,Vogel:1975}
forming a bound muon-nucleus state, a muonic atom.
The $\mu^{-}$ in the muonic atom will typically reach the lowest 1S energy level.
The weak interaction of the bound
muon with the nucleus leads to \textit{nuclear} muon capture
$\mu^{-}+(Z,A)\rightarrow\nu_{\mu}+X$
competing with muon decay.
Here $X$ is a final state consisting of a residual nucleus
and perhaps one or more protons, neutrons, gammas, etc.,
\cite{Measday:2001yr}.
In this paper ``muon capture'' refers to \textit{nuclear} muon capture.

Muon capture and neutrino-nucleus interaction are closely related
processes. The capture process probes nuclear response in the energy
range below $100\MeV$, providing a valuable validation of theoretical
models of importance to current and future neutrino experiments
\cite{Nieves:2004wx,Nieves:2011pp}.
Upcoming Mu2e \cite{Mu2e-TDR} and COMET \cite{COMET:2013} experiments
will stop intense negative muon beams in aluminum to look for charged
lepton flavor violation.  Most of the muons will undergo the nuclear
capture process.  Secondaries from the capture will create unwanted
detector rates.  In particular, protons and deuterons from capture are
highly ionising and can deaden the tracker.  It is important to
understand their rates to optimize the detectors.  There is an ongoing
joint effort between Mu2e and COMET to study muon capture on aluminum in
a dedicated AlCap experiment \cite{Litchfield:2015usa}.  The present
analysis, on the other hand, uses an opportunistic $\mu^{-}$ dataset
acquired by the TRIUMF Weak Interaction Symmetry Test (TWIST)
experiment \cite{Rodning:2001js,TWIST:2011aa}.  The same data were
previously analyzed for the electron spectrum from muon decay in orbit
(DIO) \cite{Grossheim:2009aa}.

A theoretical model to describe the yield and spectrum of protons from
muon capture was developed by Lifshitz and Singer
\cite{Lifshitz-Singer:1978, Lifshitz-Singer:1980}.
It considers both pre-equilibrium and compound-nucleus emission from the
excited nucleus with only 7\% of the yield from the pre-equilibrium phase
for muon capture on Al.
The agreement between the results of these models and data available at
that time was
reasonably good. The predicted yield for proton emission (plus any number of
neutrons) is 4.0\% per capture and the corresponding yield for deuterons is 1.2\%.
For the highest-energy part of the spectrum, the impulse approximation
is inadequate and capture on pairs of nucleons must be included \cite{Lifshitz-Singer:1988}.

Published data on the emission of charged partices from
nuclear muon capture
come from several types of experiments.
Studies with photographic emulsions
\cite{Morinaga-Fry-emulsion-capture,Kotelchuck-Tyler-emulsion-capture,Vaisenberg:1970tx}
provide information on both the spectrum and the yield of charged
particles, but are limited in their choice of the target nucleus.
Another spectrum measurement with an active target was done by
stopping negative muons in a silicon detector
\cite{Sobottka-Wills-si-capture}.  Measurements with external
scintillator calorimeters were done for multiple elemental targets,
see \cite{Budyashov:1971vv,Krane:1979wt} and references therein.
However, they were only able to detect a small high energy tail of the
charged particle spectrum, with the bulk of the spectrum being below
the detector threshold.  (Note that the proton spectrum in, for example,
\cite{Sobottka-Wills-si-capture} peaks at $2.5\MeV$, and a proton at
this energy has a range of only dozens of microns in material.)
Another class of experiments measures radioisotope production after
irradiating a target with negative muons
\cite{Wyttenbach:1978rp,Heusser:1973fa}.  Those experiments allow to
deduce the total amount of charged particles emitted in the capture
process, but do not provide any information on their spectra, and can
not differentiate reaction channels that produce the same final state
nucleus. (For example emission of $p$ and $n$ versus emission of a
deuteron.)
The present work is the first observation of the \emph{momentum}
spectra of charged particles from muon capture in a magnetic
spectrometer, and it is subject to different systematic uncertainties
than calorimetric \emph{energy} spectra measurements in earlier
experiments and in AlCap.

In the absence of data about most of the spectrum of protons from muon
capture on Al, the experimental spectrum
of charged particles from muon capture on Si
\cite{Sobottka-Wills-si-capture} was parameterized by
E.V. Hungerford \cite{Hungerford:1999} for the MECO experiment
\cite{MECO}.  MECO did not proceed past the design stage, but the same
parameterization was later used by Mu2e and COMET for their studies.
From the (sometimes contradictory) literature Hungerford estimated the
per capture rate for protons to be 10\%, and approximated the spectrum
as
\begin{equation}
   W(T) = A(1-T_{th}/T)^{\alpha}e^{-T/T_0}
  \label{eq:MECO}
\end{equation}
for kinetic energy of the proton ${T<8\MeV}$, and as a pure exponent in kinetic
energy with different slopes below and above $20\MeV$ \cite{Hungerford:1999}.
Although some versions of GEANT4 \cite{Agostinelli:2002hh} include code to
generate charged particle emission after muon capture, the ``MECO
spectrum'' frequently has been used as a practical solution for
simulations of the energy spectrum of protons from muon capture to
compare with data or for design purposes.

\section{Experimental Setup}

\begin{figure}[!hbt]
  \includegraphics[width=3.4in]{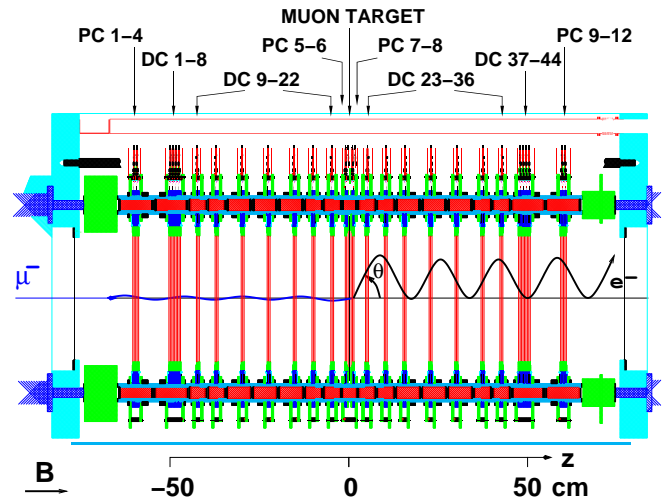}
  \caption{\label{f:TWIST_spectrometer}(color online)
    A cross section of the TWIST detector array,
    including an example $\mu^{-}$ decay event.
  }
\end{figure}

The TWIST detector (Fig. \ref{f:TWIST_spectrometer}) has been
described in detail in earlier publications, see
\cite{Henderson:2004zz,Bueno:2011}.  In this section,
components that are of particular interest for this analysis will be discussed.
The muon beamline was configured to transport a negative beam
of approximately $29\MeVc{}$ with about 1\% momentum bite.
The incoming beam contained cloud muons (negative muons generated in the proximity of the
production target) at a rate of 80~Hz.
A thin plastic scintillator at the upstream end of the spectrometer
provided a trigger signal, with a threshold chosen to be mostly
insensitive to electrons that dominated the beam flux.
The trigger was unbiased for muon decay or capture products.

The muons were then transported into the
center of the detector and stopped in a $71\um{}$ thick 99.999\% pure
Al target.
The muon range was adjusted using a gas degrader controlled by a feedback loop.
Two stacks of 22 high-precision planar drift-chambers (DCs)
were located upstream and downstream of the target.
In addition, a total of 8 multi-wire proportional chambers
(PCs) were placed at the very upstream and downstream ends of the
detector to support the event reconstruction by providing timing
information.  The target was surrounded by another 4 PCs (PC5,6
upstream and PC7,8 downstream) to enable the measurement of the
stopping position of each individual muon. The target itself served
as the cathode foil of the two innermost PCs; thus the gas volumes
surrounding the target were sensitive. A DC consisted of 80 sense wires at 4 mm
pitch surrounded by 6\um{} thick aluminized Mylar cathode foils
separated by 4~mm, filled with dimethyl ether at
atmospheric pressure. Most DCs were paired into modules of two
(so-called u and v modules) with the central foil shared. A relative
rotation of the wire planes by 90 degrees allowed for the
reconstruction of the position of a hit in the perpendicular
plane. The PCs were of similar design, but their wire planes were
equipped with wires at 2~mm pitch and a ``faster'' gas (80:20 mixture
of $\text{CF}_4$ and isobutane at atmospheric pressure) was chosen.
Charge collection time in the PCs was faster than 100~ns, while in the DCs it could exceed $1\us$.
In addition to the leading edge time, the time-over-threshold was
recorded as an estimate of the size of the signal, approximately
proportional to the energy deposit.
The gains of PCs 5 and 6 were reduced to avoid saturation of the
signal for slow muons; this reduced their detection efficiency for
electrons drastically.
However, these PCs could then be used to select muons that stopped in the Al
target rather than the PC6 gas.
The space between the chambers was  filled with a (97:3) mixture of helium and nitrogen.
The complete detector was
contained in a superconducting solenoid magnet, providing a highly
uniform field of 2 Tesla.

\section{\label{sec:simulation}Monte Carlo simulation}

The Monte Carlo simulation (MC) of the experiment that was developed for
previous TWIST measurements was adapted for this analysis.
It contains a detailed description of the geometry of the detector and
the response of its components to particles traversing the detector
\cite{Bueno:2011,TWIST:2011aa}.
The MC is implemented in GEANT 3.21 \cite{GEANT:1994}.  Its output
is in the same format as the detector data and is processed by
the same reconstruction program.
The simulation was carefully validated by extensive comparisons with
data \cite{MacDonald:2008xf,TWIST:2011aa}.  The accuracy of the
geometric description and the amount of material in the detector are
of particular importance to the current analysis.
Earlier TWIST studies did not cover the passage of protons and heavier
particles emitted in the muon capture process through detector
material.  We use ``out of the box'' GEANT 3.21 description of such
processes for this analysis, and assess a resulting systematic
uncertainty based on agreement with our data, as described in
Sec.~\ref{sec:momentum-range-systematic}.

The simulation starts with a muon entering the detector. (Beam
electrons and pile-up muons are also simulated with appropriate
rates.)  The muon is tracked through the detector with all its GEANT~3
interactions enabled.  Typically it ranges out and stops in detector
material, at which point a custom routine is called to produce a
capture proton, deuteron, etc., or a decay in orbit electron.  The
daughter particle is subsequently tracked through the detector with
all its interactions enabled.

The muon beam momentum and momentum bite affect the muon stopping
distribution inside the target.
That distribution is expected to have a significant effect on the
spectra of capture protons and deuterons exiting the target, because the
$70\um$ foil is thick compared to the range of a few-MeV proton in
aluminum.
A procedure based on matching MC distribution of last wire plane
hit by the incoming muon to data \cite{TWIST:2011aa}
was used to tune $\mu^{-}$ beam parameters in the simulation.
The GEANT3 code does not include the physics of the muon atomic cascade, which
affects the last hit plane distribution due to electromagnetic secondaries escaping
the target. A GEANT4 version 10.2  MC
\cite{Allison:2016lfl,Allison:2006ve,Agostinelli:2002hh}
was used to estimate the effect
and derive the corrections of $-39.4$~keV/c to MC beam momentum and
$-37.8$~keV/c to the momentum bite width.
Applying the correction shifted the mean muon stopping position by
$(-1.5 \pm 0.1)~\mu$m, and the final uncertainty on the matching
between data and MC is ${}\pm2\um$ \cite{TWIST:2011aa}.
The muon stopping distribution is further discussed in
Secs.~\ref{sec:cuts} and \ref{sec:systematics}.

Simulated samples of protons and deuterons from muon capture are only
used in this analysis to obtain the detector response function
(defined in Sec.~\ref{sec:unfolding}).  Therefore, a particular
distribution of simulated protons or deuterons in momentum is not
important as long as it covers a sufficient kinematic range.  An
exponential distribution in kinetic energy $f(E_k)=\exp(-E_k/(5\MeV))$
was used for both types of particles.  The analysis also uses samples
of electrons from muon decay in orbit (DIO), and triton and alpha
particles from muon capture.  The shape of the DIO distribution is
taken from \cite{Grossheim:2009aa}, and shapes of triton and alpha
distributions are extracted from a GEANT4 version 10.2p03 simulation
of muon capture on aluminum that used the GEANT4 precompound model
\cite{Quesada:2011}.  Technically the default constructed
\texttt{muMinusCaptureAtRest} process for the negative muon is
replaced in our simulations with a new \texttt{G4MuonMinusCapture}
instance that is given a \texttt{G4MuMinusCapturePrecompound}
argument.  The precompound model is expected to be more accurate
(\cite{Wright:2015xia}, Sec.~4.1) than the Bertini model used in
that GEANT4 version by default.

\section{Analysis}

The objective of the analysis is to determine detector-independent
normalized spectra for protons and deuterons from muon capture.
The approach is to select a sample of events where a trigger muon
stopped in the target and produced a delayed track, and reconstruct
these tracks.  Positive tracks are protons and heavier particles
emitted after muon nuclear capture, and are the main subject of this
analysis.
Reconstructed distributions of positive tracks from data
are corrected for detector efficiency and resolution effects, which
are extracted from MC, as described in Sec.~\ref{sec:unfolding}.
Negative tracks are due to DIO electrons, and are used to determine
the absolute normalization of the result,
Sec.~\ref{sec:normalization}.

The same event selection and reconstruction is applied to real
detector data and MC.  A single exception in treating data and MC in
the same way is not using a cross talk removal algorithm on MC events,
because electronic cross talk was not simulated.  This is further
discussed in Sec.~\ref{sec:systematics}.

\subsection{\label{sec:cuts}Event selection}

\begin{figure}[htp]
  \includegraphics[width=3.4in]{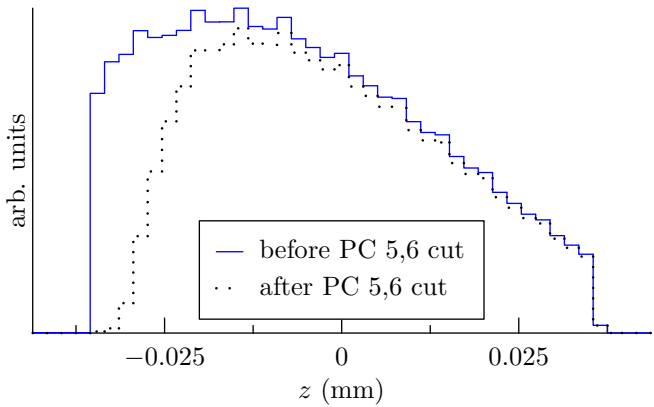}
  \caption{\label{f:muon_stopping}(color online)
    Simulated distribution of muons stopping in the Al target before (solid)
    and after (dotted) the cut to eliminate stops in PC6 gas and wires.
  }
\end{figure}

The data set considered for this analysis contains 57M triggered
events.  After removal of most crosstalk hits, the hits are grouped by
time.
The prompt time window included hits belonging to the incident muon.
A separate window starting at 400 ns included hits corresponding to
muon decay or capture products. The start of the delayed time window
was a compromise between the time required to collect all ionization
from the DC cells (up to 1000 ns) and the effective muon lifetime in Al
(864 ns \cite{Grossheim:2009aa}).
Requiring wire plane hits up to the stopping target and no hits
downstream of the target in the time window corresponding to the
trigger time, and demanding the radial position of the stopped muon to
be within 2.5~cm of the detector center, yields 22M events.  Most of
those event are due to muons stopping in the target, but that
signature is also consistent with stops in gas or wires of PC6, and
such stops comprise about 9\% the sample at this selection stage.
Out of target stops are suppressed by a cut that uses hit
time-over-threshold measurements in PC5 and PC6 as proxy for muon
energy deposition \cite{Bueno:2011,Bueno-thesis}.  The cut suppresses
out of target stops to below 0.5\%, which is negligible for the
present analysis.  It also carves the muon stopping distribution
inside the target as shown by the MC simulation in Fig.~\ref{f:muon_stopping}.
(The quasiperiodic bump structure in the figure, as well as
in Figs.~\ref{fig:sys-stopdist} and \ref{f:stopdist_pactres} below,
is an artifact caused by
an interplay of 3 different discretizations: single
precision floating point numbers in GEANT 3, packing of
real numbers into integers in TWIST data format, and the
bin size of the final histogram.)
The number of events passing the cut is 18M.

Further selection identifies events with charged particles from muon
capture or decay.
For this analysis, only the downstream chambers were used for measuring
decay or capture tracks to avoid complications from the reduced gain of PCs 5 and 6.
Also, hits from downstream decay or capture tracks are not affected by leftover
gas ionization from the incoming muon.
A delayed hit is required in one of downstream PCs 7--12, and beam pile-up
as identified by PC1--PC4 is vetoed in the delayed time window.
Then timing cuts on the delayed time window with respect to the trigger
particle and possible pile-up particles are applied.
This selection leaves 2.4M
``downstream candidate'' events.  The efficiency of cuts up to this
point is the same for DIO and reconstructable capture tracks, which
reduces systematic uncertainty on the normalization.

\subsection{Reconstruction of positively charged tracks from muon capture}

Sophisticated track finding and fitting software had been developed
to reconstruct muon decay tracks in the TWIST detector
\cite{TWIST:2011aa}.  It was modified for this analysis to optimize
its performance for proton tracks from muon nuclear capture.  Protons
have higher momentum than Michel positrons, and some are not radially
contained in the detector.  This leads to many tracks having fewer
hits for track finding compared to the positron case.  Low energy
protons range out in the detector material, and also have only a few
hits.  These effects motivated the inclusion of PC hit information in
the track finding code for the present analysis.

For the original TWIST analysis the positron trajectories were first fit to the
wire centers of the hits, and then the track was refined using drift times.
Because our drift time correction was only valid for relativistic particles,
we chose to use only the wire center fits, as this gave adequate resolution.
Energy loss and multiple scattering in the detector materials are
incorporated in TWIST fits \cite{TWIST:2011aa}.
For this work the values were changed to ones appropriate for protons.

Fiducial cuts are made to include only tracks that can be
reconstructed reliably with adequate resolution. Thus tracks must have
$0.5<cos(\theta)<0.98$, $p_t>11.9\MeVc$, and $p_z>28.4\MeVc$.
Here $\theta$ is the angle between the detector axis pointing
downstream and the initial momentum direction of the particle.
The transverse distance between the position of the stopped muon and
the extrapolation of the track to the target plane is required to be
less than $1.5$~cm.  The number of events with positive tracks passing
all the cuts is 22.3k.

\begin{figure}[htp]
  \subfloat{\includegraphics[width=3.4in]{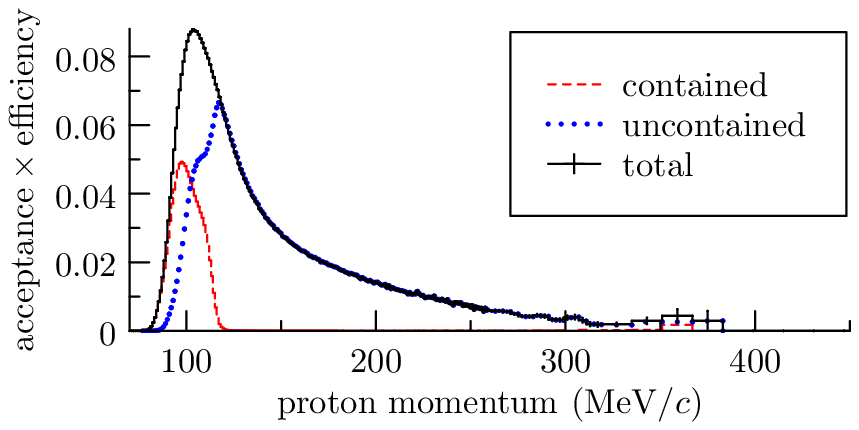}}\\
  \subfloat{\includegraphics[width=3.4in]{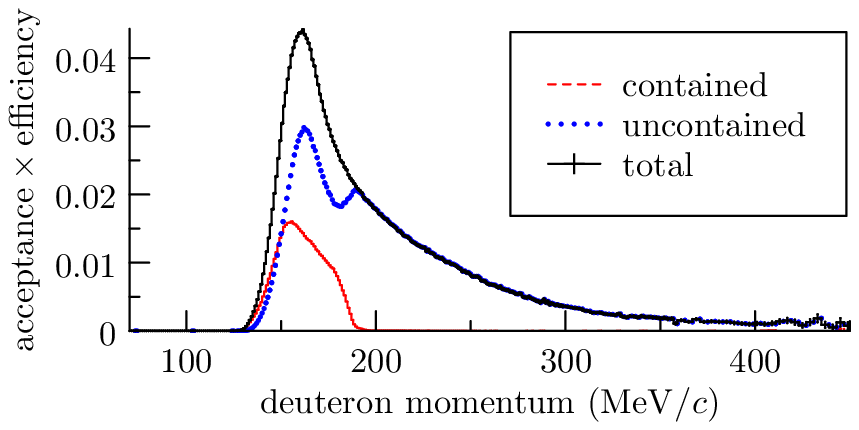}}
  \caption{\label{f:acceff}(color online)
    Reconstruction efficiency times acceptance vs momentum
    for protons (upper plot) and deuterons (lower plot)
    estimated from MC.
    The denominator includes all tracks emitted by
    muons captured in the stopping target.  Contained and uncontained
    subsamples are discussed in Sec.~\ref{sec:pid}.
  }
\end{figure}

The RMS momentum resolution of reconstructed proton tracks is better
than 8\% between 100\MeVc{} and 200\MeVc{}.  For deuterons it is
better than 15\% between 150\MeVc{} and 300\MeVc.
The product of reconstruction efficiency and acceptance for protons
and deuterons based on MC is shown in solid lines in Fig.~\ref{f:acceff}.

\subsection{\label{sec:pid}Separation of protons and deuterons}

\begin{figure*}[htp]
  \centering
  \subfloat[Data]{\label{f:pid-data}\includegraphics[width=0.51\textwidth]{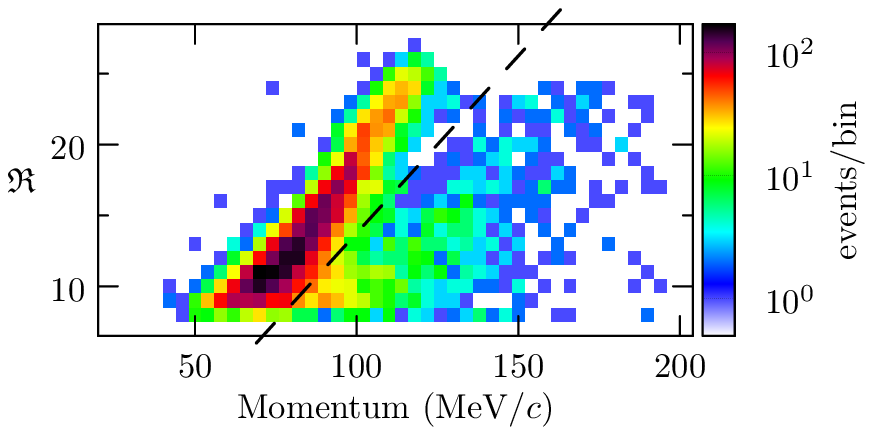}}
  \\
  \subfloat[MC proton]{\label{f:pid-mcproton}\includegraphics[width=0.49\textwidth]{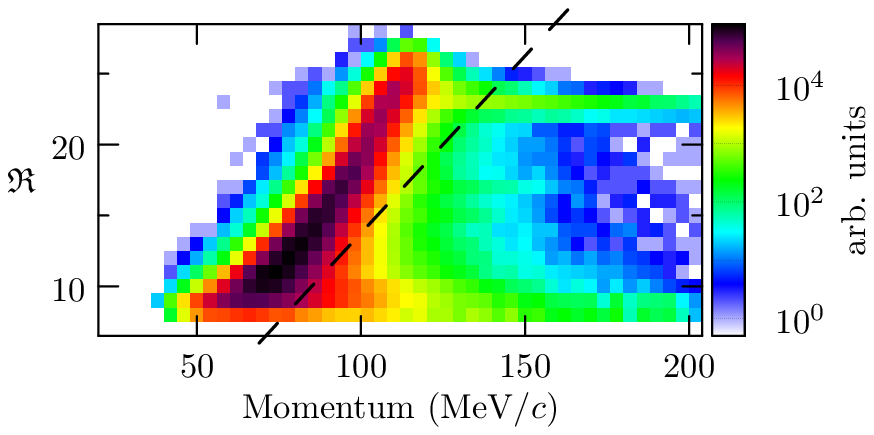}\hspace*{0.5em}}
  \subfloat[MC deuteron]{\label{f:pid-mcdeuteron}\hspace*{0.5em}\includegraphics[width=0.49\textwidth]{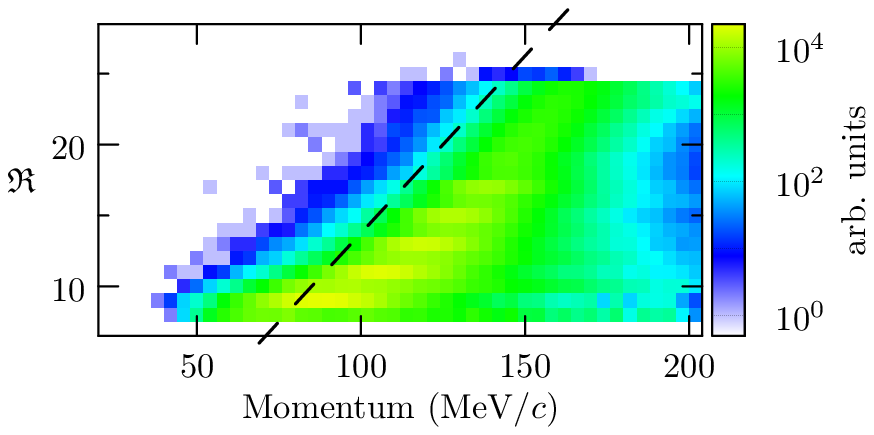}}
  \caption{\label{f:pid}(color online)
     Range $\mathfrak{R}$ vs reconstructed momentum spectra for contained tracks in data and MC.
     The dashed line is a guide to the eye to indicate proton-deuteron
     separation; it is not used in the analysis.
  }
\end{figure*}

The probability of events being protons or deuterons
is based on the difference of range in material vs
momentum relationship between protons and deuterons.  A subset of
events where the track ranges out in the detector material (contained
sample) is identified by requiring that the helical extrapolation of
the track starting at the target does not exit the instrumented part
of the detector radially, and that the range of DC planes hit by the
particle ends at or before the last downstream DC.  The cut was
based on just DC chambers because of a higher noise level observed in
the outer PCs.
The range of planes hit by the particle was defined to start at the
stopping target and include all planes with hits on the fitted track,
as well as adjacent downstream planes with contiguous hits in the
track time window.  The adjacent planes were included to account for
the fact that our tracking drops hits at the end of particle range,
because in that region the trajectory is dominated by random
scattering and the hits do not contribute information about the track
initial kinematic.
In the TWIST planar detector geometry the amount of material traversed
by a particle that crosses a detector layer is proportional to
$1/|\cos(\theta)|$.  Therefore the number of planes (DCs and target
PCs) traversed by a track was divided by $|\cos(\theta)|$ and used as the
track range variable $\mathfrak{R}$.
A distribution of track range vs momentum in data is shown in
Fig.~\ref{f:pid-data}.
MC generated plots for protons and deuterons
separately are shown in Figs. \ref{f:pid-mcproton} and
\ref{f:pid-mcdeuteron}.

Tracks that are not contained provide information about the momentum
distribution of events and are combined with the contained sample
in the likelihood function used in the analysis, as described in
Sec.~\ref{sec:unfolding}.

\subsection{\label{sec:normalization}DIO Normalization}

Both the electron spectrum and TWIST detector response to it are
known with high precision \cite{Grossheim:2009aa}, which allows to use
reconstructed negative tracks to determine the number of muon decays
in the selected sample of events.
The ratio of muon capture events to DIO in Al has been measured to be
0.609:0.391, with an uncertainty of 0.001 \cite{Suzuki:1987}, therefore
a per-capture normalization for the positive track sample can be
determined from reconstructed negative tracks.

Only downstream muon decays are counted in this analysis.  This provides
almost 0.7M DIO events for the normalization sample, rendering the
statistical uncertainty negligible.  The sample is known to be
unpolarized \cite{Grossheim:2009aa}.

The advantage of using DIO tracks for normalization is that
efficiencies of cuts that deal with muon pre-selection or pile-up
particles are identical for muon capture and decay events, providing
data-to-data and MC-to-MC cancellations in the normalization, without
requiring to match absolute cut efficiencies in data and MC.

%%%%%%%%%%%%%%%%%%%%%%%%%%%%%%%%%%%%%%%%%%%%%%%%%%%%%%%%%%%%%%%%
\subsection{\label{sec:unfolding}Unfolding}

The extraction of the desired physics result, ``truth level'' momentum
spectra $f_{\eta}(p)$ ($\eta=\{\text{proton},\text{deuteron}\}$), from
experimentally measured quantities that are smeared by the detector
resolution is called unfolding.
In this study the data observables are the bin
contents $d_i$ in the exclusive channels (contained and uncontained),
where reconstructed bin index $i$ runs over all bins in the range vs
momentum ``contained'' and momentum spectrum ``uncontained''
histograms.
The detector response $R_{\eta,i}(p)$ is the expectation value of the
number of reconstructed events in bin $i$ for a single proton (or
deuteron) emitted in the nuclear capture in the target with true
momentum $p$.  It describes all the detector effects: acceptance,
efficiency, and resolution---but is independent of the physics spectra
that are being measured.  In practice the detector response is
approximated by a matrix $R_{\eta,ij}$ obtained by binning the true
momentum variable axis:
$\int{R_{\eta,i}(p)f_{\eta}(p)\,dp}\longrightarrow\sum_{j}R_{\eta,ij}f_{\eta,j}$.
The response matrix $R_{\eta,ij}$ can be determined from Monte-Carlo
simulation.  If the bin size for the true momentum variable is much
smaller than the momentum resolution, the result does not depend on a
particular momentum spectrum used in the simulation---the only
requirement is that the simulation reaches sufficient statistics in
all of the phase space important for the measurement.  This study used
$1\MeVc$ true momentum bin size.

The expected values of the data observables can be written as
\begin{equation}
  \mu_i = N_{\text{captures}} \sum_{\eta} \sum_{j} R_{\eta,ij} f_{\eta,j}  + b_i
  ,
  \label{eq:predicted}
\end{equation}
where $b_i$ represent the background.
Because many bins in the data sample contain only a few entries, the
Poisson statistics must be taken into account.   A maximum likelihood
estimator for $f_{\eta,j}$ could be formed by maximizing
\begin{equation}
\label{eq:Poisson}
\log{\mathcal{L}}(d|\mu\{f\})  = \sum_{i} ( d_i \log\mu_i - \mu_i )
\end{equation}
for the measured set of $d_i$.  However the unfolding problem is known to
be ill-posed: truth level spectra that are significantly different
from each other can map into detector distributions that have only
infinitesimally small differences
\cite{Blobel:1984ku,Cowan1998,Cowan:2002in,Prosper:2011zz}.  The best
possible unbiased solution of an unfolding problem would have an
unacceptably large variance even if that solution saturated the
minimal variance bound \cite{Cowan1998}.
It has been shown that approximate solutions to unfolding problems can
be obtained by using a regularization procedure
\cite{Tikhonov1963a,Tikhonov1963b,Phillips1962}, which reduces the
variance of the result at the price of introducing a bias.
Regularized unfolding can be performed by maximizing a combination of
the log likelihood of data and a regularization functional
$S\{f_{\eta}\}$
\begin{equation}
\label{eq:regularization}
{\mathcal{F}} = \log{\mathcal{L}}(d|\mu\{f\}) + \alpha S\{f\}
,
\end{equation}
where $\alpha$ is the regularization parameter.  A widely used
Tikhonov regularization imposes a ``smoothness'' requirement on the
spectrum by penalizing the second derivative of the solution:
\begin{equation}
\label{eq:reg-tikhonov}
S_{\text{Tikhonov}} = -\sum_{\eta} \sum_j (-f_{\eta,j} + 2f_{\eta,j+1} - f_{\eta,j+2})^2
.
\end{equation}
It therefore biases the result towards a linear function.
Another well established regularization, the maximum entropy (or
``MaxEnt'') approach \cite{Cowan1998}, is based on the entropy of a
probability distribution \cite{Shannon:1948zz}:
\begin{equation}
\label{eq:reg-maxent}
S_{\text{MaxEnt}} = -\sum_{\eta} \sum_j q_{\eta,j} \ln(q_{\eta,j}), \qquad q_{\eta,j}\equiv f_{\eta,j}/\sum_k f_{\eta,k}
.
\end{equation}
It biases the result towards a constant.
A useful feature of the MaxEnt approach is that it guarantees that the
result will be positive everywhere, as is required for particle
emission spectra.

Our initial attempts using a maximum likelihood fit to data with
standard regularization procedures consistently
produced unphysical rising behavior of the result at the high
momentum end of the unfolding range, no matter where the end of the
range was defined.  There were two factors that caused this:  one was
the contribution of the overflow region, i.e. the part of the spectrum
above the unfolding cutoff.  While the spectrum $f_{\eta}(p)$
is falling at high momenta,
its value just above a cutoff is similar to the value just below,
leading to the last few bins in the unfolding range having similar
contribution to the reconstructed spectrum to the first few bins above
the range.  The second factor was the bias from the unfolding
regularization term.
The high momentum region has small data statistics, and is therefore
more susceptible to the bias.  This effect can be mitigated by
choosing a regularization term that biases towards the ``correct''
physics distribution.
The $f_{\eta}(p)$ functions are sharply falling---the spectra are
known to be exponential in kinetic energy for large energies.  The
Tikhonov regularization term biases towards a straight line, and
MaxEnt towards a constant; none of these is a good model for the high
energy tail of the distribution.  A cross-entropy regularization term
\cite{Cowan1998} can be used to bias towards a fixed ``reference
distribution'', but in our case the exponential slopes are unknown
parameters that we want to measure.

We addressed both of those issues by utilizing a novel unfolding
method \cite{Gaponenko:2019hpx}.
In this approach $f_{\eta}(p)$ are arbitrary functions below an
unfolding cutoff, and physically motivated exponentials in kinetic
energy above the cutoff, with parameters of the exponents included in
the fit.  The continuation of the solution with a parametrized
function outside of the unfolding region provides a correct handling
of the overflow region contribution.  Another key idea is to impose
regularization not on the complete physics spectrum, but on just the
``arbitrary function'' part of it, with the overall exponential
behavior factored out.  Then a regularization term that biases its
spectrum to a constant will bias the final physics spectrum towards
the desired exponential shape.  Specifically, we represent
\begin{widetext}
\begin{equation}
\label{eq:spectrum-parameterization}
   f_{\eta}(p) = A_{\eta} \frac{p}{\sqrt{p^2+m_{\eta}^2c^2}}\exp\{-\gamma_{\eta}T_{\eta}(p)\}
   \times
   \begin{cases}
     1 + \phi_{\eta}(p) & \text{for $p_{\eta,\text{min}} < p \le p_{\eta,u}$}, \\
     1 & \text{for $p > p_{\eta,u}$},
   \end{cases}
\end{equation}
\end{widetext}
where $p_{\eta,\text{min}}$ and $p_{\eta,u}$ determine the limit
of the unfolding region for particle type $\eta$, $m_{\eta}$ is the
mass of the particle and $T_{\eta}(p)$ its kinetic energy, $A_{\eta}$
and $\gamma_{\eta}$ are parameters pertaining to the exponential
behavior of the spectrum, and $\phi_{\eta}(p)$ is an arbitrary
function to be determined from the unfolding.
The regularization term has the form (\ref{eq:reg-tikhonov}) or
(\ref{eq:reg-maxent}) but now acts on $1+\phi$ instead of $f$:
\begin{widetext}
\begin{equation}
\label{eq:reg-tikhonov-final}
\tilde{S}_{\text{Tikhonov}} = -\sum_{\eta} \sum_j (-\phi_{\eta,j} + 2\phi_{\eta,j+1} - \phi_{\eta,j+2})^2
,
\end{equation}
\begin{equation}
\label{eq:reg-maxent-final}
\tilde{S}_{\text{MaxEnt}} = -\sum_{\eta} \sum_j \tilde{q}_{\eta,j} \ln(\tilde{q}_{\eta,j}),
\qquad \tilde{q}_{\eta,j}\equiv (1+ \phi_{\eta,j})/\sum_k (1+\phi_{\eta,k})
.
\end{equation}
\end{widetext}
We used the MaxEnt term to extract the central value and most
systematic uncertainties, and Tikhonov term to evaluate the
uncertainty related to the regularization itself.

The functions $\phi_{\eta}$ are approximated by linear
combinations of cubic basis splines $B_l$ (B-splines) \cite{Boor1978b} on their
unfolding intervals
\begin{equation}
  \label{eq:basis-splines}
  \phi_{\eta}(p) = \sum_{l}^{n_{\eta}} w_{\eta,l} B_{\eta,l}(p),
  \qquad p_{\eta,\text{min}}<p\le p_{\eta,u}
.
\end{equation}
Here $w_{\eta,l}$ are the spline coefficients determined from the
unfolding process.  We require that the resulting spectrum has a
continuous second derivative, or
$\phi_{\eta}(p_{\eta,u})=\phi'_{\eta}(p_{\eta,u})=\phi''_{\eta}(p_{\eta,u})=0$,
which is provided by having a single-fold spline knot at the endpoint
$p_{\eta,u}$.  There are no continuity constraints at
$p_{\eta,\text{min}}$, therefore a 4-fold knot is used at that point
to support the most general cubic spline shape.

The values
$p_{\text{proton},\text{min}}=80\MeVc$
and
$p_{\text{deuteron},\text{min}}=130\MeVc$
are set by the turn-on of the respective acceptance times efficiency curves,
see Fig.~\ref{f:acceff}.   The transition point between unfolding and exponential
fit regions
$p_{\text{proton},u}=230\MeVc$
and
$p_{\text{deuteron},u}=200\MeVc$,
as well as the number and
position of intermediate knots, were optimized based on unfolding of
multiple, statistically independent mixed samples of simulated
protons, deuterons, and the backgrounds described below.  The mixed
samples were designed to have statistics similar to the statistics of
the actual data sample.  The Appendix
provides more information.

The unfolding accounted for two sources of background $b_i$ in
Eq.~(\ref{eq:predicted}): contribution of heavier than deuteron particles
emitted in muon capture (tritons and alphas), and DIO electrons
misreconstructed as positive tracks. The shape of each of the two
backgrounds was taken from the simulation described in
Sec.~\ref{sec:simulation},
while its normalization was a free fit parameter $\beta$:
\begin{eqnarray}
b_i &=& b_{h,i} + b_{e,i}\nonumber\\
&=& \beta_{h}\,\frac{b^{\text{MC}}_{h,i}}{\sum_{j}b^{\text{MC}}_{h,j}}
  + \beta_{e}\,\frac{b^{\text{MC}}_{e,i}}{\sum_{j}b^{\text{MC}}_{e,j}}
.
\nonumber
\end{eqnarray}

The regularization strength $\alpha$ in Eq.~(\ref{eq:regularization})
should be chosen to provide an optimal balance between the variance
and the bias of the result.  The method used in this study was
inspired by the L-curve approach \cite{Hansen:1992,Hansen:1993},
and is the following.  For
a given $\alpha$ the maximization of
$\tilde{\mathcal{F}} = \log{\mathcal{L}} + \alpha \tilde{S}$
 yields particular values of
$\log{\mathcal{L}}$ and $\tilde{S}$.  We consider a parametric curve
$(\log\log{\mathcal{L}}(\alpha),\log{|\tilde{S}(\alpha)|})$, and use the $\alpha$
corresponding to the point of the maximum curvature of that curve as
the optimal solution.  This choice of L-curve variables for the
log-likelihood fit and Tikhonov regularization corresponds to the
classical L-curve defined for a $\chi^2$ unfolding \cite{Hansen:1992}.
For the
MaxEnt regularization we explored several functions of $\tilde{S}$ and chose
$\log|\tilde{S}|$ based on the tests with Monte-Carlo samples.

Both data and simulation sample sizes contribute to the statistical
uncertainty of the result.  To estimate the data contribution, the
central values of $f_{\text{proton}}(p)$, $f_{\text{deuteron}}(p)$,
and the backgrounds were determined first by performing unfolding on
the actual data sample.  Those values were used to calculate
expectation values for each data bin $\mu_i$ using
Eq.~(\ref{eq:predicted}).  Then ``pseudodata'' contained and
uncontained histograms were produced by sampling an appropriate
Poisson distribution for each bin, and the whole unfolding procedure,
including a new choice of the optimal $\alpha$, was performed on that
pseudodata sample.  The procedure was repeated for 25 statistically
independent pseudodata samples, and the variance of $f_{\eta}$ was
computed from pseudodata unfolding results.  The contribution of
simulation statistics was estimated by splitting a simulation sample
into parts and performing unfolding on real data using only one part
of that MC sample at a time, and looking at the variation of the
result.  The sizes of MC samples were confirmed to be sufficiently
large, so that the statistical uncertainty of the result is dominated
by the real data statistics.

\begin{figure*}[htp]
  \includegraphics{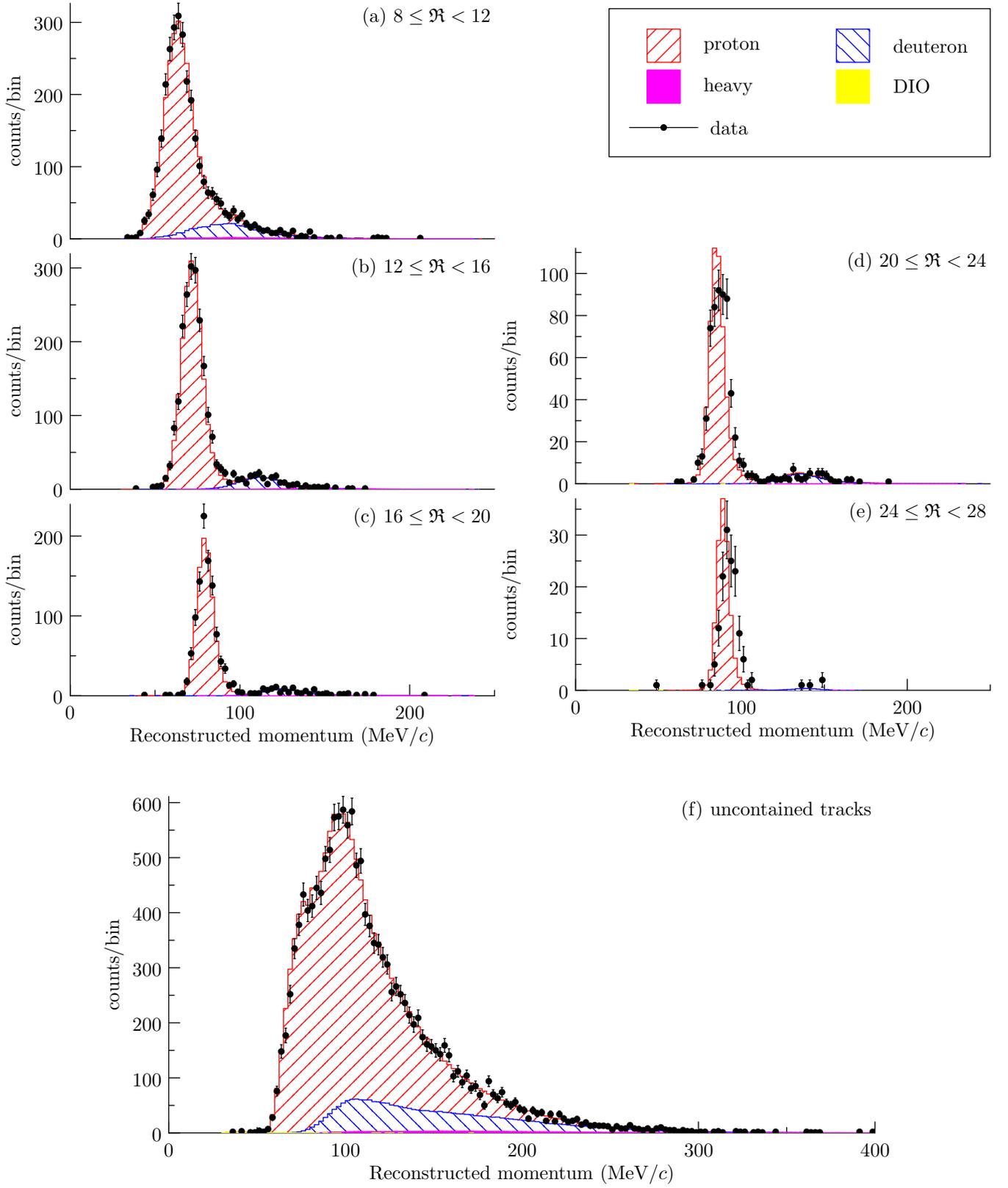}
  \caption{\label{f:fitSlices}(color online)
    Reconstructed momentum in slices of track range, and
    stacked simulation distributions after fit to data.}
\end{figure*}

The momentum spectrum of reconstructed data tracks is shown in
Fig.~\ref{f:fitSlices}.  Panels (a)--(e) show the spectrum of the
contained tracks
in slices of  the
track range variable $\mathfrak{R}$ (defined in Sec.~\ref{sec:pid}).
The panel (f) shows the spectrum of uncontained tracks.
The fitted values of $\mu_i$ are also shown in these plots.  Fit component
yields are 19.1k protons, 2.9k deuterons, 250 heavier than deuteron
particles, and 12 misreconstructed DIO events for the 22.3k sample of
data tracks.
The fit agreement with data is excellent for the uncontained spectrum,
and is good for most bins of the contained spectrum. There is a
discrepancy for longer range tracks, clearly visible in
Fig.~\ref{f:fitSlices}(e).  The peak in the reconstructed MC track
momentum distribution is shifted down compared to the data peak by
about 3\%.  This is further discussed in
Sec.~\ref{sec:momentum-range-systematic}.

Section \ref{sec:results} below shows unfolded $f_{\eta}(p)$, and
the Appendix provides more technical details on the fit results.

\section{\label{sec:systematics}Systematic uncertainties}

A discrepancy between the simulation of an effect and its actual
impact on data would introduce a systematic error in the result.
Systematic uncertainties are estimates of the size of such errors.  In
this study they are typically evaluated by modifying a parameter in simulation
and measuring the effect on the resulting yields and spectra by
performing the complete unfolding procedure with the modified sample.
Another source of systematic error is the unfolding bias.
It is estimated by performing unfolding of data with baseline MC
samples while modifying some of the settings in the procedure.  The
rest of this section provides details on individual uncertainties,
and Table \ref{t:systematics} gives a summary of their effects on the
measured yield of protons and deuterons.
Figures \ref{f:protonYield}, \ref{f:deuteronYield} in the Results
section show the momentum dependent total uncertainty of the unfolded
spectra, which includes a quadratic sum of all the systematic
uncertainties discussed in Secs.~\ref{sec:mustop}--\ref{sec:method}
below.

Among systematic uncertainties the most important ones for proton are
cross talk (below 200\MeVc) and method bias (above 200\MeVc).  For
deuteron, the energy loss systematic dominates in most of the momentum
range below 300\MeVc{}, and cross talk and heavy particle background
uncertainties become dominant above 300\MeVc{}.

\subsection{\label{sec:mustop}Muon stopping position}

\begin{figure}[!hbt]
  \includegraphics[width=3.4in]{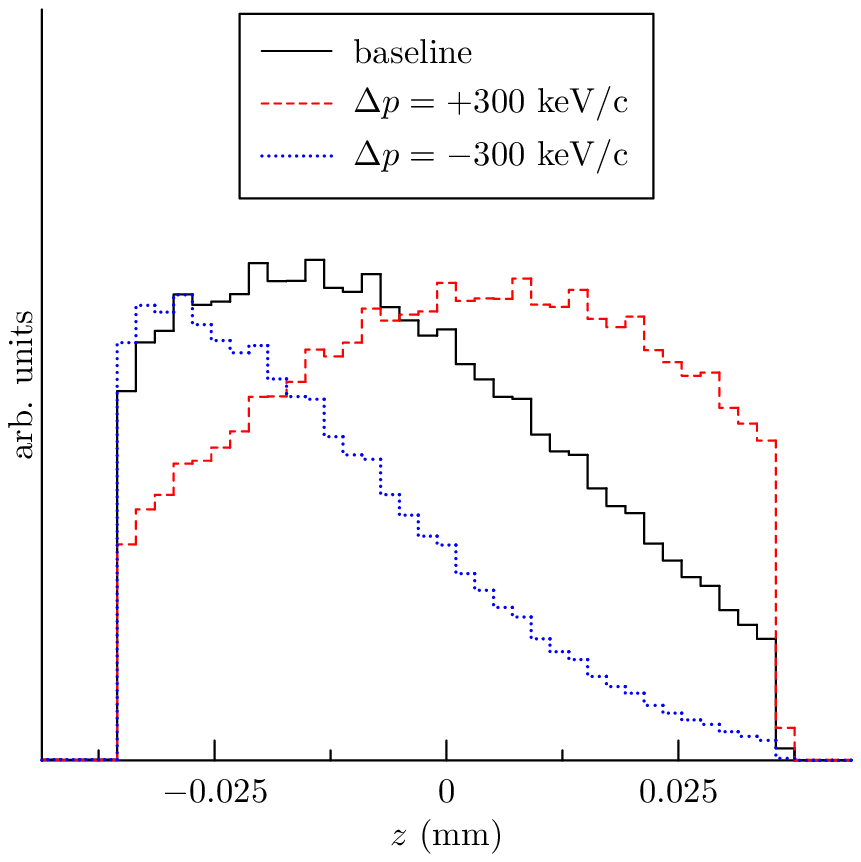}
  \caption{\label{fig:sys-stopdist}(color online)
    Simulated distribution of muons stopping in the Al target for the baseline
    and modified beam momentum samples.
  }
\end{figure}

The muon beam momentum in the baseline simulation was tuned as
described in Sec.~\ref{sec:simulation}, resulting in a $\pm2\um$
uncertainty on the stopped muon position.  Modified simulation
samples were produced by changing the beam momentum by $\pm300\keVc$.
The resulting shifts in the stopped muon distribution are shown
in Fig.~\ref{fig:sys-stopdist}.  The $300\keVc$ momentum
modification shifted the peak position by about 10 times the
uncertainty.  The stopping position systematic uncertainty was evaluated by
taking the difference between data unfolding results with the
$+300\keVc$ sample and $-300\keVc$ sample, and scaling it down by a
factor of 20.  Although the simulation samples were produced by
modifying the beam momentum, this systematic uncertainty also covers
other effects that shift the stopping distribution, such as
differences in the amount of material in the muon path between data
and MC and simulation of muon $dE/dx$.

\subsection{Carving of the muon stopping distribution by the PC5,6 cut}

\begin{figure}[htp]
  \includegraphics[width=3.4in]{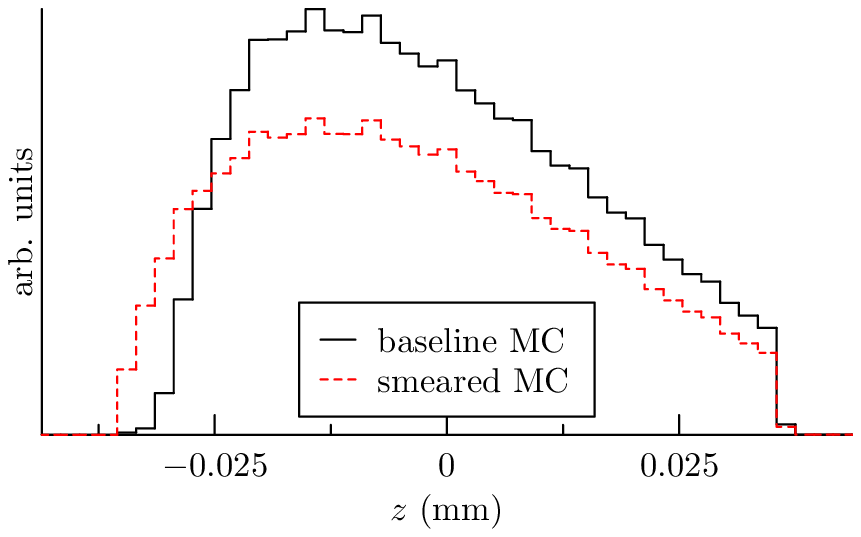}
  \caption{\label{f:stopdist_pactres}(color online)
    Simulated distribution of muons stopping in the Al target for
    the baseline and smeared simulation
    after the PC5,6 cut.
  }
\end{figure}

The PC5,6 cut that suppresses out of target stops changes the shape of
the muon stopping distribution for the accepted event sample, see
Fig.~\ref{f:muon_stopping} in Sec.~\ref{sec:cuts}.  The effect of
the cut depends on the resolution of the cut variable, which is the
time over threshold measurement in the wire chambers.  A comparison of
time over threshold spectra for data and MC showed that simulated
distributions were sharper than data.  To evaluate the corresponding
systematic effect, time over threshold values were smeared in
simulated samples to make MC shapes match those in data, and full
analysis chain re-done with the smearing applied.  The effect of the
smearing on the muon stopping distribution is shown in
Fig.~\ref{f:stopdist_pactres}.  We treat the difference between unfolding
results for the baseline and smeared analyses as the systematic
uncertainty corresponding to the chamber hit time over threshold
modeling.

\subsection{Crosstalk}

% The details are in 20170726 talk.

% The effect is linear across 0.02, 0.05 and 0.10 flag drop fractions:
% the resulting uncertainty is the same for the corresponding unfolding
% given its SF.

Electronic crosstalk was a minor effect for minimum ionizing positron
tracks from muon decay \cite{TWIST:2011aa}, and was not implemented in
the simulation.  Instead, an algorithm to flag crosstalk hits in data
was developed.  The algorithm uses the fact that crosstalk hits are
delayed with respect to the real hit that induced them, and have
smaller time over threshold values than most real hits.  Hits flagged
as crosstalk were ignored by track reconstruction.

Crosstalk is more important for heavily ionizing proton and deuteron
tracks.  To evaluate its effect in the present analysis, data
reconstruction and unfolding were re-done after ``un-flagging'' at
random a fraction of identified crosstalk hits.  An increased amount
of crosstalk at tracking input reduces the number of reconstructed
tracks and changes the unfolded spectrum.  This effect is linear in
the ``un-flagging'' fraction range from 0 to at least 0.1.  Therefore
the difference between the unfolding result for an ``un-flagging''
setting that doubles the amount of residual crosstalk hits and the
baseline result provides an estimate of the systematic uncertainty
caused by imperfect crosstalk removal.

\begin{figure}[htp]
  \includegraphics[width=3.4in]{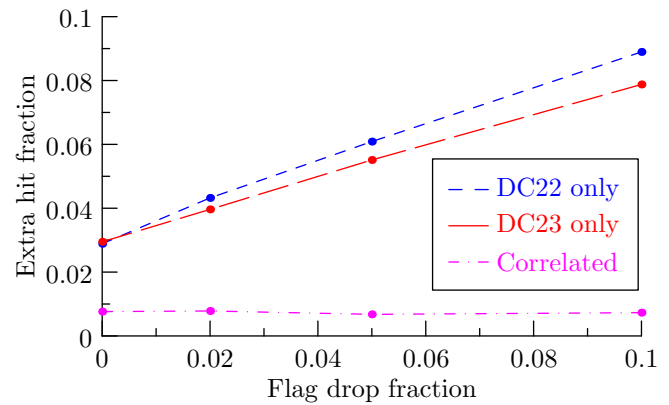}
  \caption{\label{f:xt-scaling}(color online)
    Dependence of the observed correlated and
    uncorrelated extra hits in data on the crosstalk ``un-flagging''
    fraction.  }
\end{figure}

To determine the correct ``un-flagging'' setting to measure the
systematic, a sample of data events with positive tracks having
$\theta<40^{\circ}$ was analyzed for different configurations of hit
wires in a DC plane.  A single wire hit is attributed to a real signal
from the track.  Two adjacent wires may be due to two real or one real
and one ``extra'' hit.  A track with less than a $45^{\circ}$ angle can
not cause ionization in more than 2 square gas cells, therefore if
more than 3 wires are hit there are ``extra'' hits in the plane.  Also
any configuration with non-adjacent hit wires must have ``extra''
hits.  Some of the ``extra'' hits are crosstalk, while others are
produced by extra particles in the event (such as a delta electron
from the proton track).  An extra particle can produce hits in
multiple planes, while crosstalk hits in different planes are not
correlated.  This allows to disentangle the two effects.  Figure
\ref{f:xt-scaling} shows observed rates of correlated and uncorrelated
hits in DC22,23.  As is expected, the rate of correlated hits does not
change with the ``un-flagging'' fraction setting, while the rate of
uncorrelated hits grows linearly, confirming their crosstalk origin.
The baseline rate of uncorrelated extra hits doubles for the the
``un-flagging'' fraction of 0.05, therefore this setting was used to
determine the crosstalk systematic uncertainty.

\subsection{\label{sec:momentum-range-systematic}Charged particle energy loss in detector material}

This analysis relies on the modeling of proton and deuteron energy
loss in material by out-of-the-box GEANT 3.21 \cite{GEANT:1994}.  The
discussion in section PHYS430 of the GEANT manual suggests that the
model is accurate to a few percent for low energy protons.
Reconstructed momentum spectra of contained tracks in our setup
provide an experimental handle on the quality of the energy loss
model.  There is a visible discrepancy between data and simulation in
the position of the momentum distribution peak for protons ranging out
deep in the detector, see Fig. \ref{f:fitSlices}(e).  The most
probable momentum in data is 92.4\MeVc{} while in MC it is
88.7\MeVc{}.  The about 4\% difference is consistent with the expected
accuracy of the simulation.

To evaluate the impact of the discrepancy on the final result, various
changes to the simulation code have been tried.  The original GEANT3
computed energy loss for protons was scaled by velocity dependent
factors.  In other simulation runs the amount of material in various
parts of the detector was varied.  While some of the variations
improved data-MC agreement for contained tracks with long range, most
of them broke the agreement for shorter contained or uncontained
tracks.  The variation that showed the best improvement for long
contained tracks without introducing discrepancies in other regions
was a 20\% increase in the density of the drift chamber DME gas.
Because the increased gas density also affects the muon stopping
distribution, the muon beam momentum in the simulation was adjusted to
mostly compensate for the change.  A small correction compensating for
a residual shift of the muon stopping distribution was applied to the
difference between unfolding results with the modified simulation and
the baseline to obtain the final estimate of the systematic uncertainty.

\subsection{Heavy particle background simulation}

\begin{figure*}[htp]
  \includegraphics{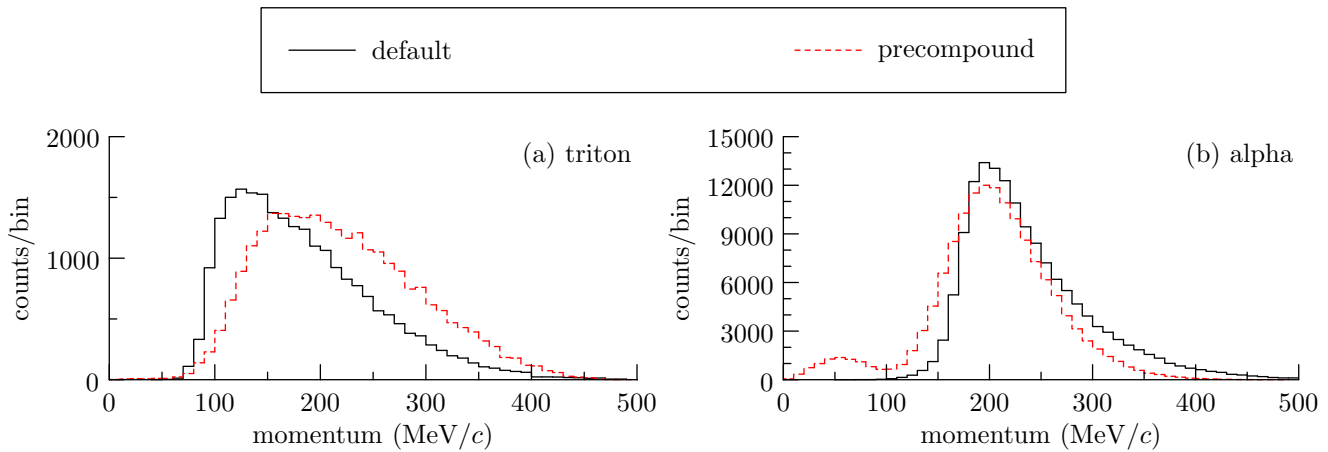}
  \caption{\label{fig:ta-models}(color online)
    Generator level triton and alpha
    particle momentum spectra for $10^7$ muon nuclear captures on
    aluminum simulated with GEANT4 precompound and Bertini models.}
\end{figure*}

The data unfolding fit included triton and alpha particles from the
nuclear capture as a single background component with fixed shape and
floating normalization.  The shapes of triton and alpha distributions
and their relative yields for the baseline fit were taken from the
GEANT4 precompound model as described in Sec.~\ref{sec:simulation}.
The Bertini Cascade as implemented in GEANT4 \cite{Wright:2015xia}
provided an alternative model for the heavy particle spectra.  The
difference between the models is illustrated in Fig.~\ref{fig:ta-models}.
The systematic uncertainty was estimated as the difference between the
unfolding results when using baseline and alternative inputs for the
heavy particle background.

Protons are well separated from heavier particles by track range and momentum,
so the proton yield is only weakly affected by the
modeling of triton and alpha emissions, as can be seen in
Table~\ref{t:systematics}.  The deuteron result, on the other hand, is
directly influenced by simulation of this background.

\subsection{\label{sec:method}Method bias}

The proton and deuteron spectra are approximated by cubic splines.  To
estimate the systematic uncertainty resulting from such approximation,
the unfolding procedure was re-done by increasing the number of
splines in the basis set by one for protons or deuterons, one
variation at a time.  The variation of the proton parameterization
changed the measured partial yield by $0.00028$ for protons and
$0.00022$ for deuterons.  The variation of the deuteron
parameterization produced a $0.00012$ change in the proton result and
$0.00013$ in the deuteron one.

The regularization term in the unfolding procedure biases the
result. To estimate the bias, Tikhonov regularization term was used
instead of the default MaxEnt form.  This resulted in a $0.00008$
change in the proton yield and $0.00005$ change in the deuteron
yield.

The quadratic sum of the pararameterization and regularization effects
is shown in Table~\ref{t:systematics} in the ``Method'' line.

\subsection{Other effects}

Deuteron breakup in detector material is a potential source of
systematic uncertainty. A dedicated MC using cross sections from
Refs. \cite{Auce:1996, Huu-Tai:2006} showed that the fraction of
deuterons breaking up in the detector was at the level of $10^{-3}$,
and thus negligible.

The background from DIO electrons misreconstructed as positive charge
tracks is included in the fit. It contributes only at lowest momenta
and contributes negligible uncertainty to the measured spectra.

The geometry of the detector and the spectrometer magnetic field are
known to a high accuracy \cite{Henderson:2004zz,Bueno:2011}, making
the uncertainy due to momentum scale calibration negligible.

The proton and deuteron spectra are normalized per muon capture as
described in Sec.~\ref{sec:normalization}.  The uncertainties due
to the knowledge of the DIO spectrum and the data sample size have a
negligible effect on the final result.

%%%%%%%%%%%%%%%%%%%%%%%%%%%%%%%%%%%%%%%%%%%%%%%%%%%%%%%%%%%%%%%%
\begin{table}
  \caption{\label{t:systematics}Statistical and systematic uncertainties on partial yield of protons and deuterons per muon capture on Al.}
\begin{ruledtabular}
\begin{tabular}{lcc}
Uncertainty & proton & deuteron \\
\hline
Crosstalk & 0.00205 & 0.00018 \\
Stopping position & 0.00010 & 0.00001 \\
Stopping distribution & 0.00049 & 0.00020 \\
Energy loss & 0.00025 & 0.00032 \\
Background & 0.00004 & 0.00036 \\
Method  & 0.00032 & 0.00026 \\
\hline
Combined systematic & 0.00215 & 0.00061 \\
Data statistics & 0.00066 & 0.00088 \\
MC statistics & 0.00006 & 0.00009 \\
\hline
Total & 0.00225 & 0.00108
\end{tabular}
\end{ruledtabular}
\end{table}

%%%%%%%%%%%%%%%%%%%%%%%%%%%%%%%%%%%%%%%%%%%%%%%%%%%%%%%%%%%%%%%%
\section{\label{sec:results}Results}

\begin{figure*}
  \includegraphics{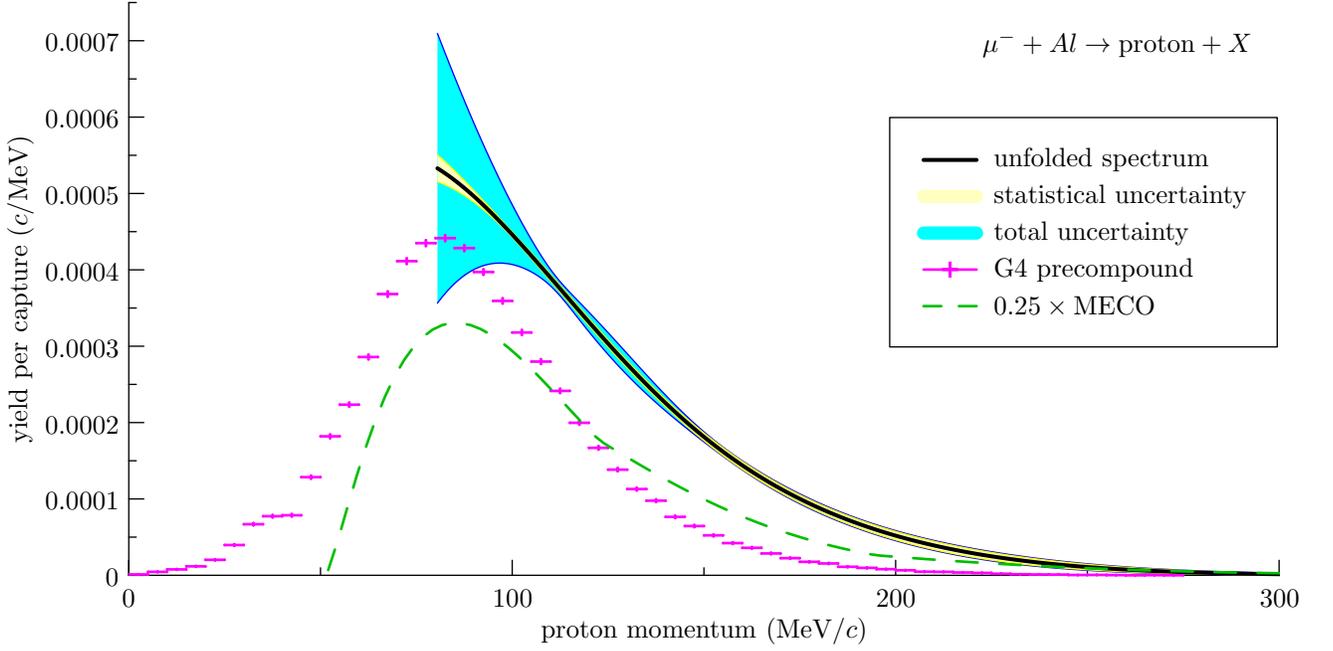}
  \caption{\label{f:protonYield}Yield of protons per muon capture vs momentum.}
\end{figure*}

\begin{figure*}
  \includegraphics{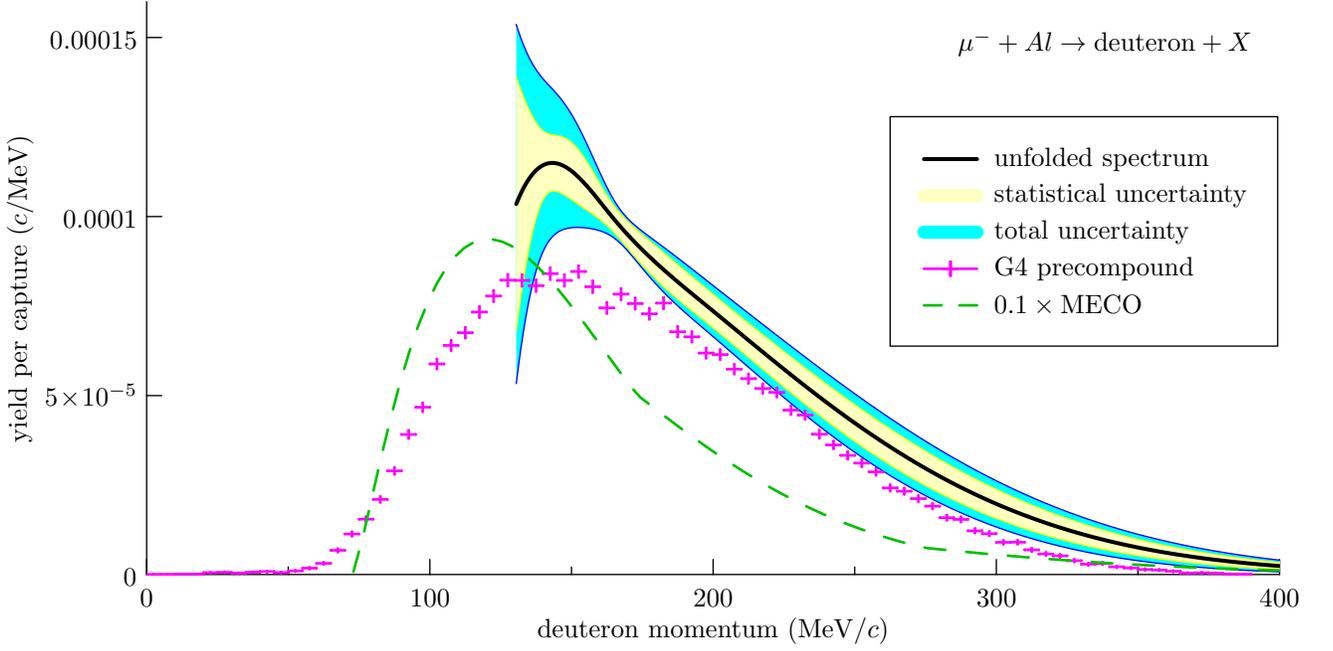}
  \caption{\label{f:deuteronYield}Yield of deuterons per muon capture vs momentum.}
\end{figure*}

Figures \ref{f:protonYield} and \ref{f:deuteronYield} show the
differential yields of protons and deuterons
per muon capture $f_\text{proton}(p)$ and $f_\text{deuteron}(p)$.
As described in Sec.~\ref{sec:unfolding}, the
unfolding splines are defined in the 80 to 230 \MeVc{} range for
protons, and 130 to 230 \MeVc{} for deuterons.  The higher momentum data
included in the fit assume exponential (in kinetic energy)
behavior, so there is no explicit upper cut off for the fit
range.  We define an effective upper cut off for this measurement as
the momentum beyond which the expected contribution of the exponential
tail to the reconstructed spectrum equals one event. This
cut off is about
300\MeVc{} for protons and 400\MeVc{} for deuterons.
The uncertainty on the proton spectrum below 150\MeVc{} is mostly
systematic, while above 150\MeVc{} it is dominated by the data sample statistics.
For deuteron the statistical and systematic contributions are of
similar size in most of the range.

The figures also show the ``MECO spectrum'' \cite{Hungerford:1999}
where deuterons are assumed to have the same spectrum shape in kinetic
energy as protons, and spectra predicted by GEANT4 version 10.2p03 with
the precompound model, as detailed in Sec.~\ref{sec:simulation}.
(The default codes for muon
capture are in poor agreement with our data and are not shown.)

Integrating the unfolded spectra while taking into account bin-to-bin
correlations, we obtain the per capture yields of
\begin{equation}
0.0322\pm0.0007(\text{stat})\pm0.0022(\text{syst})
\end{equation}
for protons above $80\MeVc$ and
\begin{equation}
0.0122\pm0.0009(\text{stat})\pm0.0006(\text{syst})
\end{equation}
for deuterons above $130\MeVc{}$.  The correlation coefficient between
the visible yields, including all statistical and systematic
uncertainties, is $-0.25$.

To extrapolate to the total yield we normalize the GEANT4 precompound
or MECO prediction to match our measured
spectrum just above the threshold, and add the integrated yield up to
our threshold to the measured yield.  We take the average between
GEANT4 precompound and MECO results as the central value, and
include half the difference in the extrapolation uncertainty.
Another, smaller, contribution to the extrapolation uncertainty is
obtained by varying the matching momentum. The resulting per capture
yields are
\begin{equation}
0.045\pm0.001(\text{stat})\pm0.003(\text{syst}) \pm 0.001(\text{extrapolation})
\end{equation}
for protons and
\begin{equation}
0.018\pm0.001(\text{stat})\pm0.001(\text{syst})\pm 0.002(\text{extrapolation})
\end{equation}
for deuterons.
These numbers are close to the corresponding theoretical
predictions of $0.040$ and $0.012$
\cite{Lifshitz-Singer:1980}.
Radioisotope yields for muon
capture on aluminum are reported in \cite{Wyttenbach:1978rp} and
\cite{Heusser:1973fa} as $0.028\pm0.004$ for the final state
consistent with $(pn)$ emission, and $0.035\pm0.008$ for $(p2n)$.
Note that the activation measurement technique can not distinguish
separate $p$ and $n$ particles from a deuteron.  The sum of the earlier
measurements $0.063\pm0.009$ compares well to our sum of extrapolated
proton and deuteron yields of
$0.062\pm0.004$.

\begin{figure*}
  \includegraphics{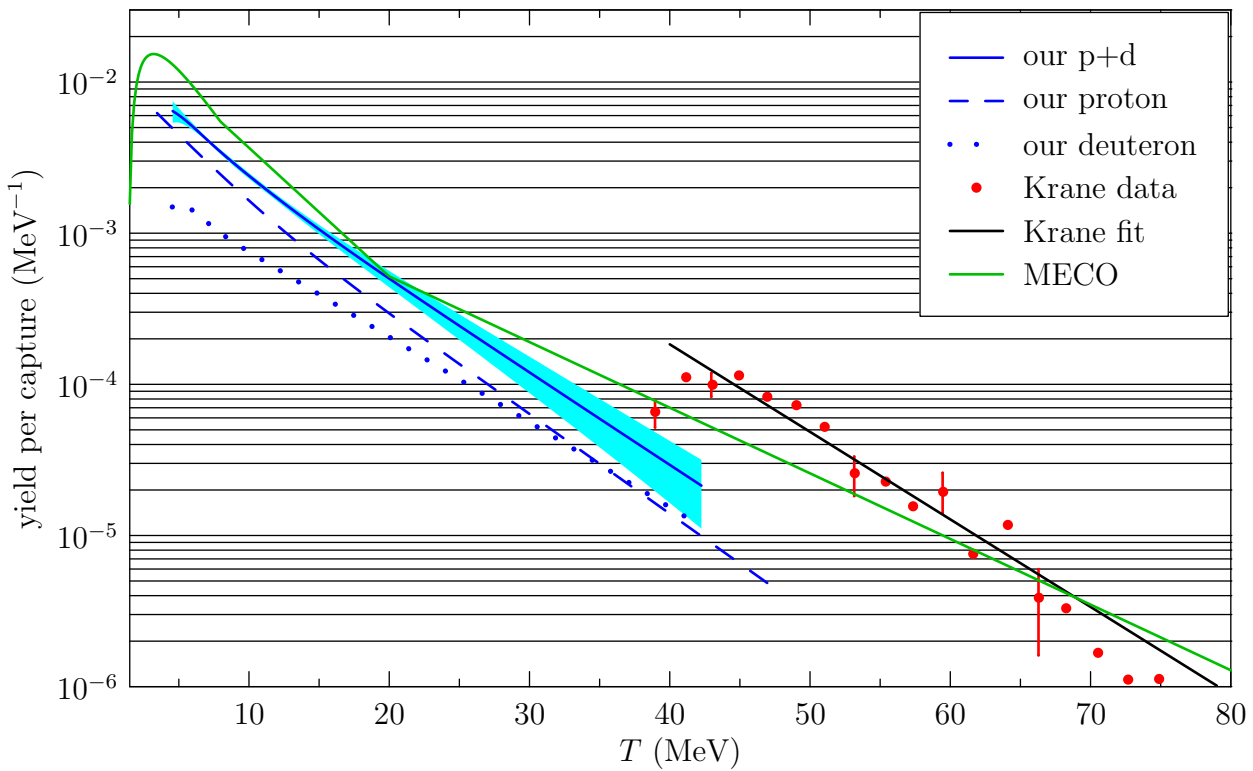}
  \caption{\label{f:krane}Yield of charged particles per muon capture vs kinetic energy.}
\end{figure*}

A measurement of the energy spectrum of charged particles from muon
capture on aluminum above $40\MeV{}$ is reported in
\cite{Krane:1979wt}.  Figure \ref{f:krane} reproduces data points and
the fit from that paper, along with our proton and deuteron results
converted to kinetic energy spectra.  While the slope of the fit in
\cite{Krane:1979wt} agrees with that of our $p+d$ curve, the
previously reported yield around $40\MeV$ is significantly higher.
Figure \ref{f:krane} also shows that the MECO spectrum
\cite{Hungerford:1999} provides,
withing the uncertainties,
an upper bound on the sum of proton
and deuteron emission.  Individual proton and deuteron spectra are
bound by the MECO curve along with the 65\% proton and 35\% deuteron
emission fractions suggested in \cite{Hungerford:1999}.

\section{Summary}
The TWIST data taken with a $\muminus$ beam incident on Al have been
analysed for positive charged particles from muon capture.
The detector was sensitive to protons with momentum above 80\MeVc{}
and deuterons above 130\MeVc{}.
These ranges
are found to contain about $70\%$ of the yield for each particle.
A precision of better than 10\% over the momentum range of
100--190\MeVc{} for protons is obtained; for deuterons of
145--250\MeVc{} the precision is better than 20\%.
The results are presented in
a format suitable for building an event generator for use in
simulations. They have immediate application to the design of
experiments searching for $\mu$ to $e$ conversion.
These spectra are of interest as input to an improved theoretical
understanding of the physics of muon capture on nuclei. They are
strongly selective of the GEANT4 precompound model over the other
presented choices.
The total per capture yields of
$0.045\pm0.003$
for protons and
$0.018\pm0.002$ for deuterons are the most precise
measurements of these quantities to date.

\begin{acknowledgments}

We thank all TWIST collaborators who contributed to detector
construction and data taking, as well as TRIUMF staff.  We thank
Konstantin Olchanski who promptly fixed any computer problems we
had.
This work was supported in part by the Natural Sciences
and Engineering Research Council of Canada, the Russian Ministry of
Science, and the U.S.A. Department of Energy.
One of the authors (Andrei Gaponenko) was supported by the resources
of the Fermi National Accelerator Laboratory (Fermilab), a
U.S. Department of Energy, Office of Science, HEP User
Facility. Fermilab is managed by Fermi Research Alliance, LLC (FRA),
acting under Contract No. DE-AC02-07CH11359.

\end{acknowledgments}

%%%%%%%%%%%%%%%%%%%%%%%%%%%%%%%%%%%%%%%%%%%%%%%%%%%%%%%%%%%%%%%%
\appendix*
\section{\label{sec:appendix-parameters}Parameterization of the unfolded spectra}

\begin{figure}[htp]
  \includegraphics[width=3.4in]{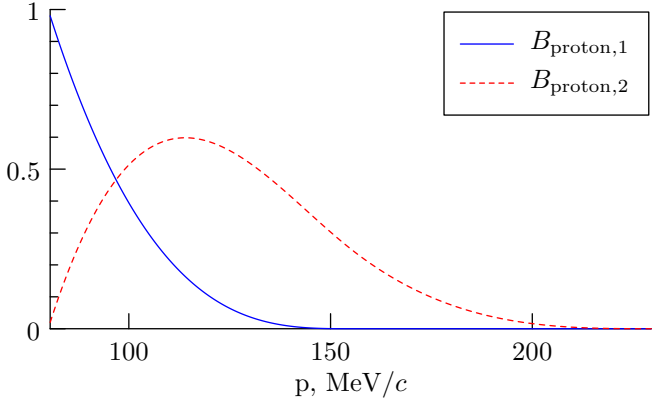}
  \caption{\label{fig:splines}(color online)
    Basis splines for the proton spectrum.
  }
\end{figure}

The notation used in the Appendix is defined in
Sec.~\ref{sec:unfolding}.  The parameterization of the measured
spectra is given by Eq.~(\ref{eq:spectrum-parameterization}), with the
arbitrary functions $\phi_{\eta}(p)$ approximated by linear
combinations of cubic B-splines \cite{Boor1978b} per
Eq.~(\ref{eq:basis-splines}).
The splines for the proton spectrum are defined by the sequence of
knots $80,80,80,80,155,230\MeVc$, resulting in a set of two basis
functions that are illustrated in Fig.~\ref{fig:splines}.  The deuteron
spline sequence is $130,130,130,130,165,200\MeVc$, similarly
resulting in two cubic splines.
The parameters of the unfolded proton and deuteron spectra are shown
in Table~\ref{tab:param-values}, and their correlations in
Table~\ref{tab:param-correlations}.
The uncertainties shown in Table \ref{tab:param-values} include
statistical and most systematic contributions.  The parameterization
shape uncertainty is not representable in terms of the fixed set of
fit parameters, and is the only contribution that is not included.

\begin{table}[htbp]
\begin{ruledtabular}
\begin{tabular}{l|cc}
Parameter & Value & Uncertainty \\
\hline
$A_{\text{proton}}$ ($c/\text{MeV}$) & $6.0\times10^{-3}$  & $2.4\times10^{-3}$ \\
$\gamma_{\text{proton}}$ ($\text{MeV}^{-1}$) & $1.5\times10^{-1}$  & $2.2\times10^{-2}$ \\
$w_{\text{proton},1}$ & $7.4\times10^{-1}$  & $6.8\times10^{-1}$ \\
$w_{\text{proton},2}$ & $5.1\times10^{-1}$  & $5.5\times10^{-1}$ \\
$A_{\text{deuteron}}$ ($c/\text{MeV}$) & $2.7\times10^{-3}$  & $5.9\times10^{-4}$ \\
$\gamma_{\text{deuteron}}$ ($\text{MeV}^{-1}$) & $1.3\times10^{-1}$  & $2.1\times10^{-2}$ \\
$w_{\text{deuteron},1}$ & $-3.6\times10^{-2}$  & $4.5\times10^{-1}$ \\
$w_{\text{deuteron},2}$ & $2.1\times10^{-1}$  & $3.5\times10^{-1}$ \\
\end{tabular}
\end{ruledtabular}
\caption{\label{tab:param-values}
  Parameters of the unfolded spectra.
}
\end{table}

\begin{table*}[htbp]
\begin{ruledtabular}
\begin{tabular}{l|cccccccc}
 & $A_{\text{proton}}$ & $\gamma_{\text{proton}}$ & $w_{\text{proton},1}$ & $w_{\text{proton},2}$ & $A_{\text{deuteron}}$ & $\gamma_{\text{deuteron}}$ & $w_{\text{deuteron},1}$ & $w_{\text{deuteron},2}$ \\
\hline
$A_{\text{proton}}$ & 1.00 & 0.96 & -0.48 & -0.84 & -0.14 & -0.26 & -0.03 & -0.21 \\
$\gamma_{\text{proton}}$ &  & 1.00 & -0.64 & -0.76 & -0.26 & -0.39 & -0.05 & -0.24 \\
$w_{\text{proton},1}$ &  &  & 1.00 & 0.06 & 0.48 & 0.63 & 0.28 & 0.34 \\
$w_{\text{proton},2}$ &  &  &  & 1.00 & 0.09 & 0.07 & -0.15 & -0.10 \\
$A_{\text{deuteron}}$ & &  &  &  & 1.00 & 0.90 & 0.23 & -0.36 \\
$\gamma_{\text{deuteron}}$ &  &  &  &  &  & 1.00 & 0.32 & -0.01 \\
$w_{\text{deuteron},1}$ &  &  &  &  &  &  & 1.00 & -0.04 \\
$w_{\text{deuteron},2}$ &  &  &  &  &  &  &  & 1.00 \\
\end{tabular}
\end{ruledtabular}
\caption{\label{tab:param-correlations}
  Correlation between parameters of the unfolded spectra.
}
\end{table*}

\bibliography{mucapture}

\end{document}